# $Au@MoS_2@WS_2$ Core-Shell Architectures: A Solution to Versatile Colloidal Suspensions of 2D Heterostructures

Jennifer G. DiStefano,[†,§] Akshay A. Murthy,[†,§] Chamille J. Lescott,[†,§] Roberto dos Reis,[†,‡] Yuan Li,[†,‡] Vinayak P. Dravid[*,†,‡,§]

*[†]Department of Materials Science and Engineering, [‡]Northwestern University Atomic and Nanoscale Characterization Experimental (NUANCE) Center, and [§]International Institute for Nanotechnology (IIN), Northwestern University, Evanston, Illinois 60208, USA*

**ABSTRACT:** For years, solution processing has provided a versatile platform to extend the applications of transition metal dichalcogenides (TMDs) beyond those achievable with traditional preparation methods. However, existing solution-based synthesis and exfoliation approaches are not compatible with complex geometries, particularly when interfacial control is desired. As a result, promising TMD structures, including $MoS_2/WS_2$ heterostructures, are barred from the rich assembly and modification opportunities possible with solution preparation. Here, we introduce a strategy that combines traditional vapor phase deposition and solution chemistry to build TMD core-shell heterostructures housed in aqueous media. We report the first synthesized TMD core-shell heterostructure, $Au@MoS_2@WS_2$, with an Au nanoparticle core and $MoS_2$ and $WS_2$ shells, and provide a means of suspending the structure in solution to allow for higher order patterning and ligand-based functionalization. High-resolution electron microscopy and Raman spectroscopy provide detailed analysis of the structure and interfaces of the core-shell heterostructures. UV-vis, dynamic light scattering, and zeta potential measurements exhibit the outstanding natural stability and monodispersity of $Au@MoS_2@WS_2$ in solution. As a proof of concept, the aqueous environment is utilized to both functionalize the core-shell heterostructures with electrostatic ligands and pattern them into desired configurations on a target substrate. This work harnesses the advantages of vapor phase preparation of nanomaterials and the functionality possible with aqueous suspension to expand future engineering and application opportunities of TMD heterostructures.

Suspension of transition metal dichalcogenides (TMDs) in solution has enabled attractive and unique engineering capabilities and spurred the development of numerous solution preparation methods. The many variations of wet chemical synthesis and exfoliation provide access to a wide array of TMD nanosheets suspended in solution.[1–7] In addition to the typical low cost and scalability of these solution preparation methods, they facilitate several powerful and unique opportunities for nanoscale control and engineering of TMDs, namely: (1) chemical functionalization; (2) hybridization with other functional nanomaterials; and (3) deposition and directed assembly on target substrates.[8] This accessibility and manipulation of TMDs possible in solution opens the door for multiple new technologies. Solution preparation of TMDs has benefited exploratory biomedical applications ranging from diagnostics to therapy.[9,10] Functionalized TMDs can serve as anti-bacterial agents[11] and chemotherapy drug-carriers,[12] and TMDs hybridized with plasmonic nanoparticles serve as a potent photothermal cancer treatment.[13] Additionally, a variety of fields including electrocatalysis,[14,15] sensing,[16,17] and battery technologies[18,19] have progressed due to functionalization and hybridization of TMDs in solution. Further, facile deposition of solution-prepared TMDs allows nanofabrication on substrates not amenable to traditional high-temperature synthesis, enabling an array of flexible devices including battery electrodes,[20] gas sensors,[21] memory devices,[22] and data-storage devices.[23]

Despite the numerous attractive possibilities enabled by solution-prepared TMDs, their utility has been curtailed by several notable limitations. Wet chemical synthesis suffers from inherently low processing temperatures, yielding lower TMD quality than vapor phase techniques, and solution exfoliation often produces flakes of limited size.[8,24] These limitations preclude many potential applications of solution-prepared material, including colloidal synthesis of $Au@MoS_2$ core-shell structures. Additionally, preparation of multiple layered materials simultaneously cannot be controllably achieved with current methods, especially if certain geometries or ordering are desired. Some methods are capable of forming $MoS_2/WS_2$ hybrids from solution but have historically lacked the precise control to form a single interface rather than randomly stacked aggregates.[25] This limits progress in areas where defined heterostructures are critical to innovate and create new functionality, such as high-performance electronics. Recently, a limited number of studies have expanded these methods to include a few select 2D heterostructures, including 1T-2H homojunctions[26] and $Bi_2Se_3/Bi_2Te_3$ heterostructures.[27] However, the existing solution preparation methods are still not conducive to the majority of heterogeneous structures. Notably, TMD heterostructures (e.g. $MoS_2/WS_2$) – which have shown great potential in fields including optoelectronics, catalysis, and bioimaging – are omitted in both their lateral and vertical geometries.[28–33] As a result, current methodologies limit the rich modification opportunities accessible for these promising materials.

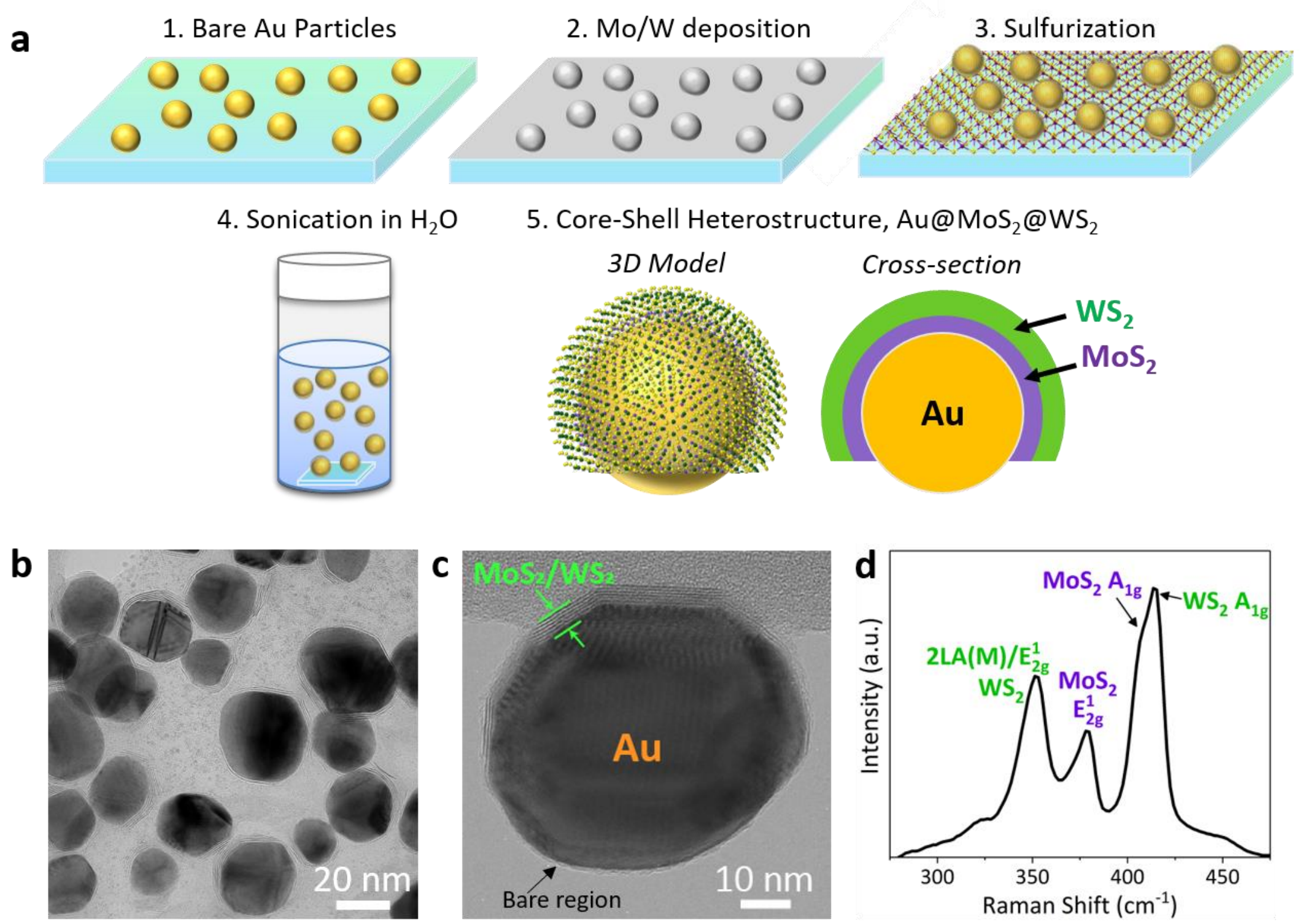

**Figure 1.** Growth schematic and structure demonstration of $Au@MoS_2@WS_2$ core-shell heterostructures. (a) Synthesis schematic depicting Mo and W metal deposition on Au nanoparticles followed by sulfurization to form $MoS_2$ and $WS_2$. Via bath sonication, the particles are then transferred to water to create an aqueous suspension of heterostructures. (b) Low magnification TEM image of core-shell heterostructures. (c) High magnification TEM of single $Au@MoS_2@WS_2$ structure where the lamellar TMD shell is visible around the Au nanoparticle core. A bare surface of the core is labeled, which is where the particle contacted the substrate during deposition. (d) Raman spectrum indicating presence of both expected TMDs.

In this study, we combine vapor phase deposition and solution chemistry to overcome this challenge and present a new TMD core-shell heterostructure suspended in water. We introduce a vapor phase thermal conversion method to produce the core-shell heterostructure $Au@MoS_2@WS_2$, where an Au nanoparticle is encapsulated by both $MoS_2$ and $WS_2$ shells. Similar structures have also been defined as core-shell-shell particles in literature. We employ the Au nanoparticle core as a vehicle to transport and suspend the surrounding $MoS_2$ and $WS_2$ layers in solution. We present data showing the unique structure of these core-shell heterostructures and high crystallinity of the TMD shells. Core-shell heterostructures in aqueous solution demonstrate the critical stability and monodispersity needed to exploit their properties, despite their lack of stabilizing ligands. Additionally, we utilize the solution environment to modify and control these nanostructures, specifically through chemical functionalization and patterning on a new substrate. This work leverages the superior material quality and control of vapor phase synthesis and the myriad of functionalization and patterning opportunities of solutions to produce colloidal suspensions of $MoS_2/WS_2$ heterostructures.

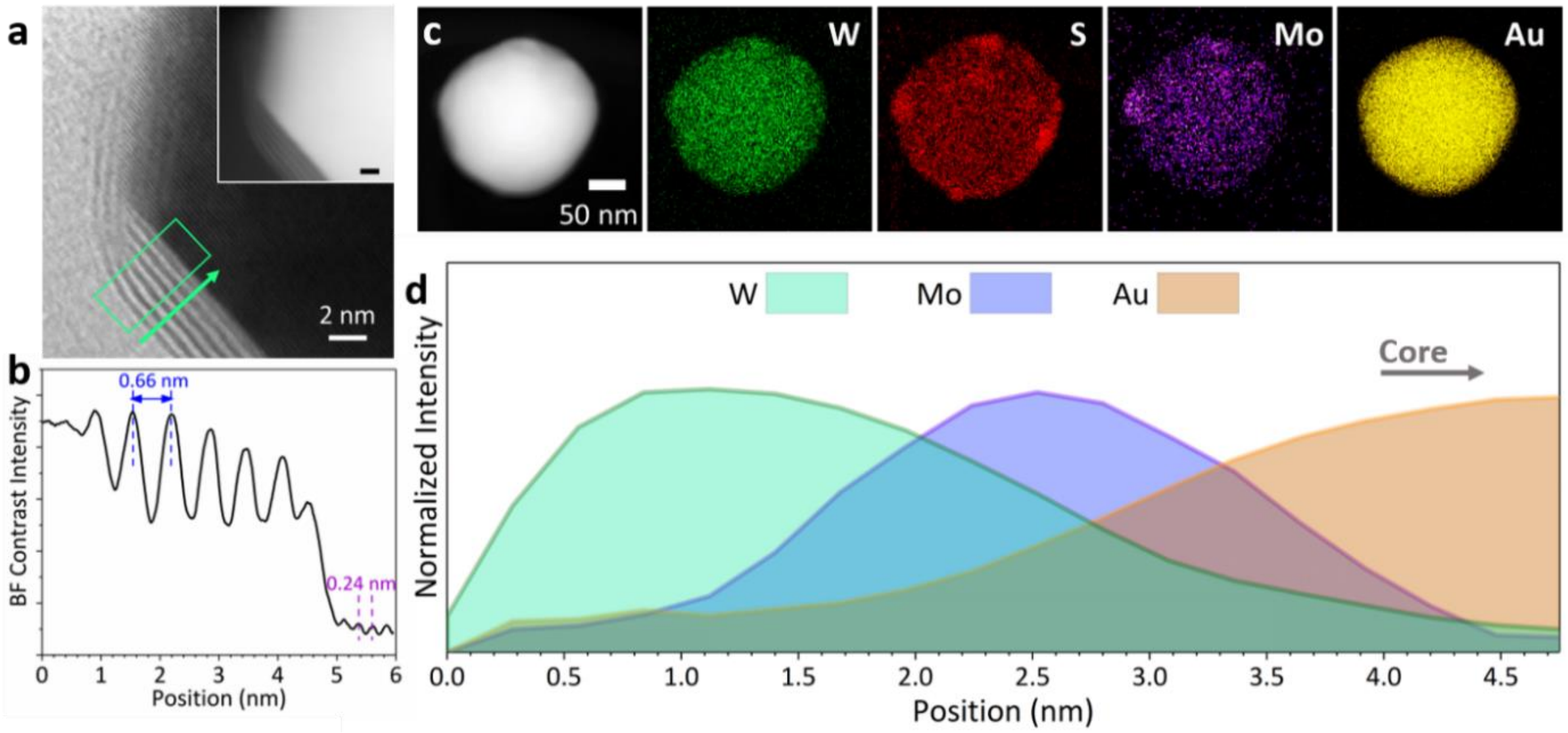

**Figure 2.** Compositional analysis of Au@$MoS_2$@$WS_2$ using STEM-EDS. (a) Annular bright field STEM image of core-shell heterostructure where the lattice fringes of the TMD shell are evident. Inset: Corresponding HAADF image (scale bar 2 nm). (b) Contrast intensity line profile taken from region of green box in (a) in direction of arrow. Distance between peaks provides the TMD layer separation of 0.66 nm and Au lattice spacing of 0.24 nm, as expected. (c) EDS maps of an Au@$MoS_2$@$WS_2$ particle showing the reference image, W, S, Mo, and Au maps, respectively. (d) EDS line profile indicating the transition from core to shell.

## RESULTS & DISCUSSION

Au@$MoS_2$@$WS_2$ core-shell heterostructures are synthesized using a modified thermally-assisted conversion technique (Figure 1a). Au nanoparticle cores are produced by evaporating 6 nm of Au onto an $SiO_2$/Si substrate followed by an 850° C anneal to form particles across the substrate. Next, a 0.5 nm layer of Mo is deposited on the prepared substrate via electron beam evaporation followed by a 0.5 nm layer of W. Finally, the substrate is sulfurized in a tube furnace to convert the Mo and W metals to $MoS_2$ image of the sulfurized layers is available in Figure S1. Figure S2 provides a scanning electron microscopy (SEM) image of the core-shell particles on their original growth substrate, with a mean diameter of 33 ± 7 nm. The substrate is then sonicated in water to transfer the particles to aqueous solution. A schematic of the final core-shell heterostructure is shown in Figure 1a.

Transmission electron microscopy (TEM) and Raman spectroscopy elucidate the structure and morphology of the core-shell particles. Figures 1b and c show high-resolution TEM (HRTEM) images of core-shell heterostructures. The low magnification image (Figure 1b) shows several encapsulated Au nanoparticles, and the TMD shell thickness is evident in Figure 1c where approximately six layers of TMD shells are visible along the top of the particle. The structures often exhibit a small bare region where the TMD shell does not cover the core, as evident along the bottom surface of the particle in Figure 1c. This exposed Au introduces additional functionality into this structure, as will be discussed later, and can be attributed to the directionality inherent in electron beam deposition and resultant lack of complete transition metal coverage. Additionally, Raman spectroscopy demonstrates the presence of each individual TMD. The Raman spectrum shown in Figure 1d exhibits the expected $A_{1g}$ (out-of-plane) and $E^1_{2g}$ (in-plane) vibration modes of both $MoS_2$ and $WS_2$ and the $WS_2$ 2LA(M) mode which overlaps with the $E_{2g}$.[34]

To further examine the TMD shell structure, scanning transmission electron microscopy (STEM) is employed. The contrast in STEM imaging in annular bright-field (ABF) and annular dark field (ADF) is nominally sensitive and proportional to atomic number Z and can therefore be used to precisely determine layer separation in the shell.[35] Figure 2b shows a contrast intensity profile from the reference ABF image in Figure 2a. In ABF images, higher Z elements appear darker due to electron absorption effects. In this structure, the brighter lines of the shell (peaks in the contrast line profile) correspond to the van der Waals gaps between TMD layers, and the darker lines correspond to the heavier elements Mo and W (valleys in the contrast line profile). The separation between layers is approximately 0.66 nm. This closely matches the expected spacing between neighboring basal planes of $MoS_2$ and $WS_2$ and confirms the Mo and W have indeed been sulfurized.[36,37] As the STEM contrast line profile extends into the core, a second set of oscillations appears corresponding to the lattice fringes of the {111} planes of Au.

Next, we employed STEM – energy dispersive X-ray spectroscopy (EDS) to gain local chemical information about the core-shells. The EDS elemental color maps shown in Figure 2c provide full-particle views of chemical composition. This data confirms coverage of the Au particle with Mo, W, and S, thus demonstrating the spatial distribution of each element is as expected. To examine the interface between materials, we conducted EDS line scans. The transition from the shell to the Au core is shown in Figure 2d, with the reference STEM image in Figure S3. Starting at the edge of the shell and moving in the direction of the core, we observe a notable increase in the W signal, corresponding to the $WS_2$ shell. As the W signal decreases, the Mo signal rises and eventually reaches its maximum, denoting the transition to $MoS_2$ as the innermost TMD. Au then becomes the most prominent element as the core is approached. More details about this analysis can be found in the Supporting Information.

While the exploration of other core-shell heterostructures is not the focus of this current work, we believe it is important to note the versatility of this core-shell synthesis method. We demonstrate the structural design opportunities by building core-shell superlattice particles, where the TMD shells alternate between $MoS_2$ and $WS_2$ (Figure S4). The demonstration of this structure exhibits the potential of this methodology to generate complex TMD heterostructures for suspension in solution moving forward.

Properties of plasmonic nanoparticles are particularly sensitive to their surrounding environment, including surface coatings and proximity to other particles. We can leverage this sensitivity of the Au nanoparticle cores to understand the ensemble structure and behavior of Au@$MoS_2$@$WS_2$ particles in aqueous solution. According to Mie theory,[38] the Au localized surface plasmon resonance (LSPR) wavelength, which can be readily measured using UV-vis spectroscopy, will red-shift as the local refractive index increases.[39] Based on this, we expect the LSPR of the core-shell particles in solution to be red-shifted compared to that of bare Au particles surrounded by only water, due to the high index of refraction of TMDs.[40,41] Indeed, the core-shell structures are observed to cause a red-shift in the Au LSPR peak location from 520 nm for bare Au nanoparticles to 566 nm for the core-shell heterostructures (Figure 3a). This shift is corroborated by previous absorbance calculations we reported on a 5-layer $MoS_2$ shell around Au particles.[42] These results indicate that the core-shell nanoparticles experience an altered dielectric environment due to $MoS_2$ and $WS_2$ shells and that ensemble encapsulation is achieved.

In addition to confirming ensemble encapsulation, the Au LSPR can also indicate the stability of a colloidal suspension. Aggregation of plasmonic particles, in our case Au, leads to coupling between neighboring particles and a notable red-shift in LSPR, the extent of which is dependent on the degree of aggregation.[43] This is often observed in UV-vis as a decrease in the expected LSPR intensity, attributed to fewer dispersed particles, and emergence of a lower energy secondary peak as the particles form aggregates.[44] Given the lack of such features in our spectra and previously stated agreement with prior calculations, the UV-vis measurements suggest the Au@$MoS_2$@$WS_2$ particles remain dispersed in water with no evidence of aggregation.

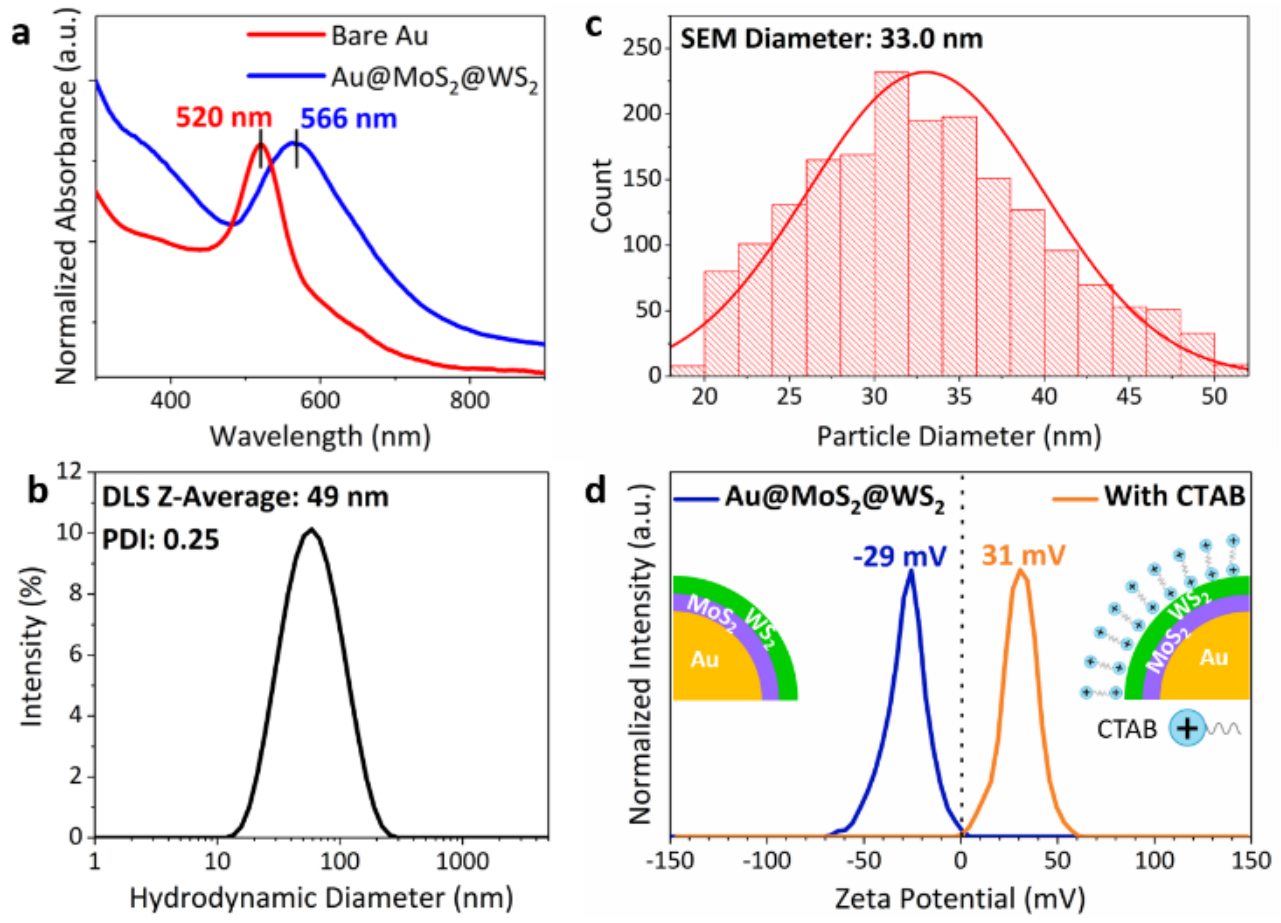

**Figure 3.** Solution characterization and functionalization of Au@$MoS_2$@$WS_2$. (a) UV-vis spectra comparing bare Au nanoparticles to core-shell heterostructures. The shift in the Au LSPR is indicative of a change in dielectric environment caused by the surrounding $MoS_2$ and $WS_2$ layers and supports that ensemble encapsulation is achieved. (b) Dynamic light scattering (DLS) measurement of the core-shell heterostructures in water reports a Z-average diameter of 49 nm and polydispersity index of 0.25. (c) SEM size distribution of core-shell heterostructures provides a mean diameter of 33 nm. This value agrees with the DLS data once the difference in measurement is accounted for, supporting that the core-shell heterostructures are monodisperse and do not aggregate in aqueous solution. (d) Zeta potential measurements of unfunctionalized Au@$MoS_2$@$WS_2$ and functionalized with CTAB. Without functionalization, the core-shell heterostructures exhibit an impressively negative zeta potential. This negative repulsion between particles in solution likely explains the stability observed in DLS. Additionally, we observe the zeta potential with CTAB to be 31 mV, indicating CTAB attachment to the core-shell surface and successful functionalization.

To further assess the stability of the Au@$MoS_2$@$WS_2$ solutions, we utilized dynamic light scattering (DLS), a hydrodynamic size measurement technique. By measuring the size of species in the solution, we can determine both the dispersity of the particles and presence of contaminants. DLS reported a Z-average diameter and polydispersity index (PDI) of 49 nm and 0.25, respectively. The size distribution from DLS is shown in Figure 3b, and SEM particle size measurements are shown in Figure 3c for reference. It is important to note that DLS measures hydrodynamic size and therefore cannot be expected to exactly match the SEM size values (see Supporting Information for detailed explanation); however, the shape of the distributions provide information about the state of the solution. The DLS data indicate that no unwanted contaminates are present in the sample, which would manifest as additional peaks (typically at larger sizes) visible in the size distribution. Similarly, formation of large Au@$MoS_2$@$WS_2$ aggregates in solution would also result in an extra peak. We can further look to the PDI value of the distribution to rule out the formation of small nanoparticle aggregates.[45] The PDI approximates the average uniformity of particles in solution on a scale from 0 to 1, where a PDI of zero represents a monodisperse solution comprised of perfectly uniform particles.[46] As such, a high PDI (>0.4) would indicate either a very broad particle size distribution or aggregation. Our reported PDI of 0.25 is well within the expected range of 0.1-0.4, considered "moderate polydispersity".[47] We expect some level of polydispersity simply because the size of the particles is not completely homogeneous, as evident by the SEM size distribution. Therefore, the combination of these two size measurement techniques indicates that the core-shell heterostructures are stable and do not aggregate in aqueous solution. This further corroborates our observations from UV-vis that the core-shell heterostructures form a monodisperse suspension.

This monodispersity of Au@$MoS_2$@$WS_2$ in aqueous solution is somewhat surprising given that no external stabilizing ligand is attached to the $WS_2$ surface of the particles. TMD sheets in aqueous solution typically require stabilization by surfactants or surface receptors to prevent aggregation.[48] To investigate the origin of stability of core-shell heterostructures, we measured the surface charge of particles in solution using zeta potential. Interestingly, we find the zeta potential to be very negative (Figure 3d), suggesting that electrostatic repulsion between particles is the stabilizing mechanism of these solutions. The zeta potential of the core-shell solutions approaches the -30 mV threshold generally indicative of a stable solution.[48] We hypothesize this large surface charge is linked to the retention of excess negative charge in few-layer TMD geometries, as other groups have reported previously.[49–51]

The natural stability and monodispersity of the Au@$MoS_2$@$WS_2$ solution makes it an ideal platform for accessing the modification and manipulation opportunities of solution preparation – functionalization, hybridization, and assembly. Chemical functionalization can provide the additional functionality necessary to make these solutions viable for future avenues such as programmable assembly and biomedical applications. We tested the potential for chemical functionalization of these particles using the electrostatic linker cetyltrimethylammonium bromide (CTAB). The large positive charge inherent to CTAB molecules allows for stabilization of particles in solution through repulsion of nearby particles and is detectable using a zeta potential measurement.[52] Figure 3d shows the zeta potential of CTAB-functionalized Au@$MoS_2$@$WS_2$ particles following dialysis, which is done to remove excess CTAB, compared to the unfunctionalized particles. The addition of CTAB causes a significant positive shift in the zeta potential, as expected, confirming successful surface functionalization.

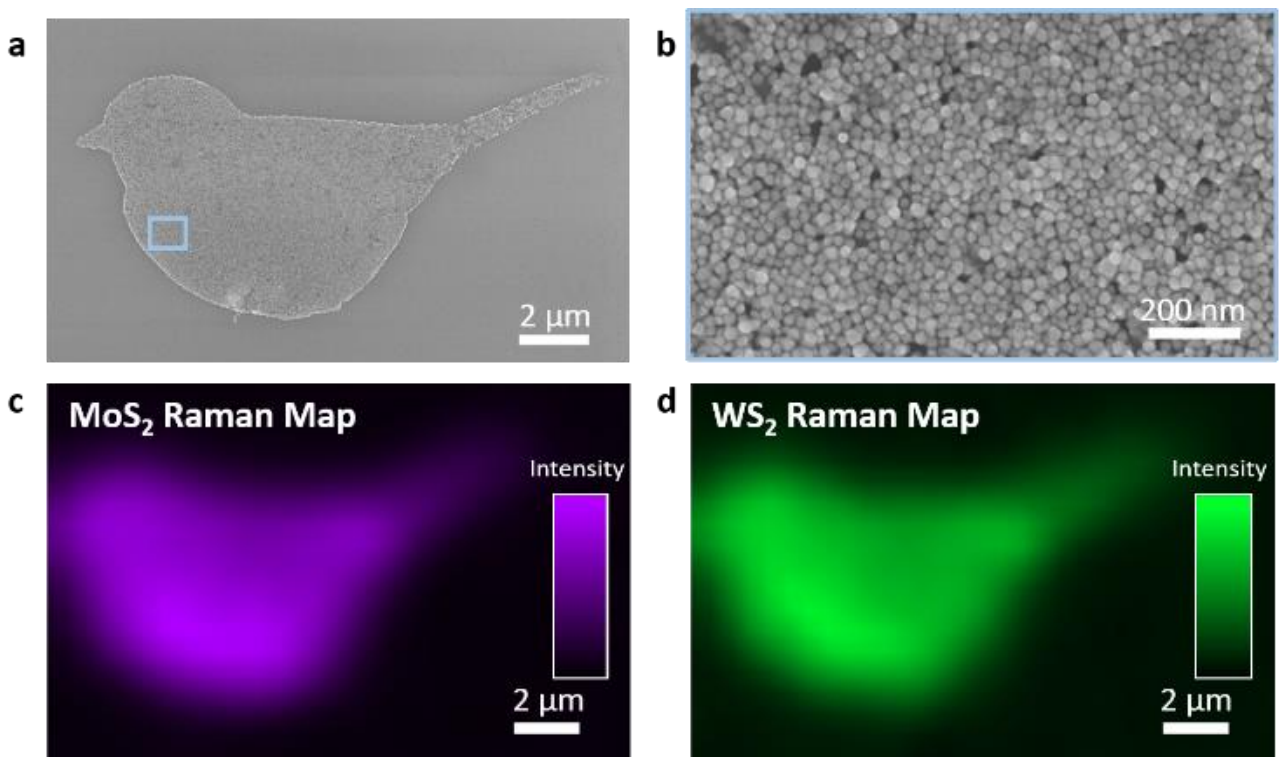

**Figure 4.** Patterning Au@$MoS_2$@$WS_2$ particles from solution. (a) SEM image of core-shell heterostructures assembled into patterns using electron-beam lithography. (b) High magnification SEM from blue box region in (a) showing nanoparticles comprising pattern. (c) $MoS_2$ Raman map of pattern. (d) $WS_2$ Raman map of pattern. Raman maps were generated using the respective $E^1_{2g}$ Raman signal for each TMD.

The ability to functionalize core-shell heterostructures offers a powerful lever to control particle chemistry and placement. For example, CTAB itself can provide a simple route to self-assemble nanoparticles. As a solution of CTAB-functionalized nanoparticles dries, capillary forces drive the nanoparticles to assemble into a monolayer or various exotic patterns, which paves a route for facile patterning of the Au@$MoS_2$@$WS_2$ system that does not rely on advanced nanofabrication processes.[53] The library of potential functionalizing molecules could be notably expanded through the inclusion of covalently-bound molecules by exploiting the thiol chemistry opportunities of the $WS_2$ outer shell or exposed Au surface. One could even imagine utilizing ligand-directed self-assembly to construct artificial crystals or site-specific patterns using Au@$MoS_2$@$WS_2$ as building blocks.

Further, particles in solution possess the distinct advantage over substrate-bound materials that they can be controllably patterned or hierarchically assembled onto a host of target substrates. This may serve as a particularly valuable approach to employ core-shell heterostructures as building blocks in a variety of future devices, where both the pattern design and type of substrate provide unique functionality. The stability and monodispersity of core-shell heterostructures in solution, which are critical prerequisites to patterning, enable us to provide a proof-of-concept demonstration that Au@$MoS_2$@$WS_2$ can indeed be controllably patterned into arbitrary configurations from solution. SEM images of core-shell heterostructures patterned using electron beam lithography are provided in Figure 4a with the corresponding high magnification image revealing the individual nanoparticles in Figure 4b. To confirm the presence and homogeneity of $MoS_2$ and $WS_2$ comprising this pattern, Raman maps of each are provided (Figure 4c and d). The Raman maps were constructed by selecting the respective $E^1_{2g}$ modes corresponding to each TMD. This proof-of-concept demonstrates promise that these core-shell heterostructures can be precisely patterned from aqueous solution onto arbitrary substrates for future optoelectronic, sensing, and flexible devices.

## SUMMARY & CONCLUSIONS

We report a new TMD core-shell synthesis route to create stable colloidal suspensions of TMD heterostructures. This work progresses beyond previous colloidal synthesis studies of TMD core-shells by enabling solution suspension of highly crystalline TMD shells and by introducing the core-shell heterostructure architecture. S/TEM analysis confirms the unusual structural and compositional makeup of Au@$MoS_2$@$WS_2$ particles. Using UV-vis, DLS, and zeta potential measurements, we find that the core-shell heterostructures are stable and monodisperse in aqueous solution due to significant negative surface charge. We demonstrate the advantages of core-shell heterostructures in solution by showing CTAB functionalization and patterning on a new substrate. However, these are only a few examples of the many functionalization, hybridization, and assembly opportunities available with solution preparation of $MoS_2$/$WS_2$ heterostructures. The ability to attach electrostatic or thiol-based molecules to these particles provides a large parameter space of chemistries that can be employed for specific applications and sophisticated patterning. Further, the particles themselves can be customized based on design needs. Not only can the type and number of TMD layers be altered, but this structure provides a "built-in" hybridization opportunity where the core material could be selected based on desired properties for even more tunability. The core material could even be easily removed to leave a standalone TMD shell, opening another avenue of exploration.[54] We believe this study provides the groundwork for extensive future manipulation and modification of core-shell heterostructures, extending their use to unique applications only achievable with colloidal suspension.

## ASSOCIATED CONTENT

**Supporting Information**. Experimental methods, further discussion on STEM-EDS and DLS results, and supplementary figures of AFM, SEM, STEM line scan reference image, superlattice HRTEM, and additional patterns. This material is available free of charge via the Internet at http://pubs.acs.org.

## AUTHOR INFORMATION

**Corresponding Author**

* Vinayak P. Dravid: v-dravid@northwestern.edu

**Author Contributions**

The manuscript was written through contributions of all authors.

## ACKNOWLEDGMENT

This material is based upon work supported by the National Science Foundation (NSF) under Grant No. DMR-1929356 and the Air Force Office of Scientific Research under award FA9550-17-1-0348. This work made use of the EPIC, Keck-II, SPID and NUFAB facilities of Northwestern University's NU*ANCE* Center, which has received support from the Soft and Hybrid Nanotechnology Experimental (SHyNE) Resource (NSF ECCS-1542205); the MRSEC program (NSF DMR-1720139) at the Materials Research Center; the International Institute for Nanotechnology (IIN); the Keck Foundation; and the State of Illinois, through the IIN. J.G.D. gratefully acknowledges support from the National Science Foundation Graduate Research Fellowship Program (NSF-GRFP). A.A.M. gratefully acknowledges support from the Ryan Fellowship and the IIN at Northwestern University. The authors thank Dr. Benjamin D. Myers and Dr. Vikas Nandwana for helpful discussions.